\documentclass[11pt,dvips,twoside,letterpaper]{article} 
\usepackage{times,fancyhdr}
\usepackage{graphicx}
\usepackage{geometry}
\usepackage{titlesec}


\makeatletter 
\def\cleardoublepage{\clearpage\if@twoside \ifodd\c@page\else%
    \hbox{}%
    \thispagestyle{empty}
    \newpage%
    \if@twocolumn\hbox{}\newpage\fi\fi\fi} 
\makeatother 

\titlelabel{\thetitle.\quad}

\def\figurename{Figure} 
\makeatletter
\renewcommand{\fnum@figure}[1]{\figurename~\thefigure.}
\makeatother

\def\tablename{Table} 
\makeatletter
\renewcommand{\fnum@table}[1]{\centering\bfseries{\tablename~\thetable.}}
\makeatother

\usepackage{amssymb}
\usepackage{amsmath}
\usepackage{amsfonts}
\usepackage{amsthm,amscd}
\usepackage{amsbsy}
\usepackage{latexsym}
\usepackage{bm}
\usepackage{url} 			
\usepackage{mathrsfs}
\newcommand{\beq}{\begin{equation}}
\newcommand{\eeq}{\end{equation}}
\newcommand{\bea}{\begin{eqnarray}}
\newcommand{\eea}{\end{eqnarray}}

\def\<{\langle}
\def\>{\rangle}

\usepackage{dcolumn}
\usepackage{bm}
\usepackage{url}
\usepackage{amssymb}
\usepackage{amsmath}

\include{epsf} 

\newcommand{\N}{{\rm I\! N}}
\newcommand{\R}{{\rm I\! R}}

\begin{document}
\title{
{\begin{flushleft}
\vskip 0.45in
{\normalsize\bfseries\textit{Chapter~}}
\end{flushleft}
\vskip 0.45in
\bfseries\scshape Semi-Markov Models in High Frequency Finance: A Review}}
\author{\bfseries\itshape Guglielmo D'Amico$^{1,}$\thanks{E-mail address: g.damico@unich.it}, Filippo Petroni$^{2}$\thanks{E-mail address: fpetroni@unica.it} 
~and Flavio Prattico$^{3,}$\thanks{E-mail address: flavio.prattico@univaq.it}\\
$^{1}$Dipartimento di Farmacia, \\
Universit\`a "G. D'Annunzio" di Chieti-Pescara, Chieti, Italy\\
$^{2}$Dipartimento di Scienze Economiche e Aziendali,\\
Universit\`a degli studi di Cagliari, Cagliari, Italy\\
$^{3}$Dipartimento Ingegneria Industriale, dell'Informazione e dell'Economia,\\ 
Universit\`a degli studi dell'Aquila, L'Aquila, Italy}
\date{}
\maketitle
\thispagestyle{empty}
\setcounter{page}{1}

\begin{abstract}
In this paper we describe three stochastic models based on a semi-Markov chains approach and its generalizations to study the high frequency price dynamics of traded stocks. The three models are: a simple semi-Markov chain model, an indexed semi-Markov chain model and a weighted indexed semi-Markov chain model. We show, through Monte Carlo simulations, that the models are able to reproduce important stylized facts of financial time series as the persistence of volatility. In particular, we analyzed high frequency data from the Italian stock market from the first of January 2007 until end of December 2010 and we apply to it the semi-Markov chain model and the indexed semi-Markov chain model. The last model, instead, is applied to data from Italian and German stock markets from  January 1, 2007 until the end of December 2010.
\end{abstract}





\section{Introduction}

Semi-Markov chains (SMC) are a wide class of stochastic processes which generalize at the same time both Markov chains and renewal processes. The main advantage of SMC is that they allow the use of whatever type of waiting time distribution for modeling the time to have a transition from one state to another. On the contrary, Markovian models have constraints on the distribution of the waiting times in the states which should be necessarily represented by memoryless distributions (exponential or geometric for continuous and discrete time cases respectively). This major flexibility has a price to pay: the parameters to be estimated are more numerous.\\
\indent SMC generalizes also non-Markovian models based on continuous time random walks extensively used in the econophysics community, see for example \cite{mai00,rab02}.   
SMC have been used to analyze financial data and to describe different problems ranging from credit rating data modeling \cite{dam05} to the pricing of options \cite{dam09,sil04} as well as for wind energy modeling which shares similar problems with the modeling of financial data such as the strong persistence of data, see \cite{dami13,dami13bis,dami13tris}.

With the financial industry becoming fully computerized, the amount of recorded data, from daily close all the way down to tick-by-tick level, has exploded. Nowadays, such tick-by-tick high-frequency data are readily available for practitioners and researchers alike \cite{gui97,pet03}. It seemed then natural trying to verify the semi-Markov hypothesis of returns on high-frequency data, see \cite{dami11b}. In \cite{dami11b} we proposed a semi-Markov model showing its ability to reproduce some stylized empirical facts such for example the absence of autocorrelations in returns and the gain/loss asymmetry. In that paper we showed also that the autocorrelation in the square of returns is higher with respect to the Markov model. Unfortunately this autocorrelation was still too small compared to the empirical one.  In order to overcome the problem of low autocorrelation, in another paper \cite{dami11a} we proposed an indexed semi-Markov model for price return. More precisely 
 we assumed that the intraday returns (up to one minute frequency) are described by a discrete time homogeneous semi-Markov process where we introduced a memory index which takes into account the periods of different volatility in the market. It is well known that the market volatility is autocorrelated, then periods of high (low) volatility may persist for long time. We made the hypothesis that the kernel of the semi-Markov process depend on which level of volatility the market is at that time. It is to be remarked that the weighted memory index is a stochastic process which do depend on the same Markov Renewal Chain to which the semi-Markov chain is associated. Then, in our model, the high autocorrelation is obtained endogenously without introducing external or latent auxiliary stochastic processes.
To improve further our previous results, in \cite{dami12}, we propose an exponentially weighted index.\\
\indent The rest of the paper is organized as follows. Section 2 describes the semi-Markov models of financial returns. Section 3 presents the results of the three different models as applied to real financial data. Section 4 discusses some concluding remarks.

\pagestyle{fancy}
\fancyhead{}
\fancyhead[EC]{Guglielmo D'Amico, Filippo Petroni and Flavio Prattico}
\fancyhead[EL,OR]{\thepage}
\fancyhead[OC]{Semi-Markov Models in High Frequency Finance: A Review}
\fancyfoot{}
\renewcommand\headrulewidth{0.5pt} 

\section{Financial Return Models}

In this section we present the models used for modeling the financial returns. Particularly we show the SCM, the indexed semi-Markov chain (ISMC) and the weighted-indexed semi-Markov chain (WISMC).

\subsection{The Semi-Markov Chain Model}
We define a SMC with values in a finite state space $E=\{1,2,...,m\}$, see for example \cite{lim01,jans06}. Let $(\Omega,\mathbf{F},P)$ be a probability space; we consider two sequences of random variables:
\begin{equation}
J_{n}:\Omega\rightarrow E\,;\,\,\,\,\,\,T_{n}:\Omega \rightarrow \N
\end{equation}
\noindent denoting the state and the time of the n-th transition of the system, respectively.\\
\indent We assume that $(J_{n},T_{n})$ is a Markov Renewal Process on the state space $E \times \N$ with kernel $Q_{ij}(t),\,\,i,j\in E, t\in \N$.\\
\indent The kernel has the following probabilistic interpretation:
\begin{equation}
\label{due}
\begin{aligned}
&P[J_{n+1}=j, T_{n+1}-T_{n}\leq t |\sigma(J_{h},T_{h}),\,h\leq t, J_{n}=i]\\
& = P[J_{n+1}=j, T_{n+1}-T_{n}\leq t |J_{n}=i]=Q_{ij}(t)
\end{aligned}
\end{equation}
and it results $p_{ij} = \mathop {\lim }\limits_{t\, \to \,\infty } Q_{ij}(t); \, i, j \in E, \, t\in \N$
where
$
{\bf P} = (p_{ij})
$
is the transition probability matrix of the embedded Markov chain $J_{n}$.\\
\indent  Furthermore, it is useful to introduce the probability to have next transition in state $j$ at time $t$ given the starting at time zero from state $i$
\begin{eqnarray}
\label{tre}
&&b_{ij}(t)=P[J_{n+1}=j, T_{n+1}-T_{n}= t |J_{n}=i] \nonumber \\
&&=\left\{
                \begin{array}{cl}
                       \ Q_{ij}(t)-Q_{ij}(t-1)  &\mbox{if $t>0$}\\
                         0  &\mbox{if $t=0$}\\
                   \end{array}
             \right.
\end{eqnarray}
\indent We define the distribution functions
\begin{equation}
\label{trebis}
H_{i}(t)=P[T_{n+1}-T_{n}\leq t |J_{n}=i]=\sum_{j\in E}Q_{ij}(t)
\end{equation}
\noindent representing the survival function in state $i$.\\
\indent The Radon-Nikodym theorem assures for the existence of a function $G_{ij}(t)$ such that
\begin{eqnarray}
\label{quattro}
&&G_{ij}(t)=P\{T_{n+1}-T_{n}\leq t|J_{n}=i, J_{n+1}=j\}\nonumber \\ 
&& =\left\{
                \begin{array}{cl}
                       \ \frac{Q_{ij}(t)}{p_{ij}}  &\mbox{if $p_{ij}\neq 0$}\\
                         1  &\mbox{if $p_{ij}=0$}\\
                   \end{array}
             \right.
\end{eqnarray}
\indent It denotes the waiting time distribution function in state $i$ given that, with next transition, the process will be in the state $j$. The sojourn time distribution $G_{ij}(\cdot)$ can be any distribution function. We recover the discrete time Markov chain when the $G_{ij}(\cdot)$ are all geometrically  distributed.\\
\indent It is possible to define the SMC $Z(t)$ as
\begin{equation}
\label{cinque}
Z(t)=J_{N(t)},\,\,\,\forall t\in \N
\end{equation}
\noindent where $N(t)=\sup\{n\in \N: T_{n}\leq t\}$. Then $Z(t)$ represents the state of the system for each waiting time. We denote the transition probabilities of the SMC by $\phi_{ij}(t)=P[Z(t)=j|Z(0)=i]$. They satisfy the following evolution equation:
\begin{equation}
\label{sei}
\phi_{ij}(t)=\delta_{ij}(1-H_{i}(t))+\sum_{k\in E}\sum_{\tau =1}^{t}b_{ik}(\tau)\phi_{kj}(t-\tau).
\end{equation}
\indent To solve equation $(\ref{sei})$ there are well known algorithms in the SMC literature \cite{bar04, jans06}.

\indent At this point we introduce the discrete backward recurrence time process $B(t)$ linked to the SMC. For each time $t\in \N$ we define the following stochastic process:
\begin{equation}
\label{diciannove}
   B(t)=t-T_{N(t)}.
\end{equation}
\indent If the semi-Markov chain $Z(t)$ indicates the state of the system at time $t$, $B(t)$ indicates the time since the last jump.
\begin{figure}[h]
\label{traiettoria}
\centering
\includegraphics[width=8cm]{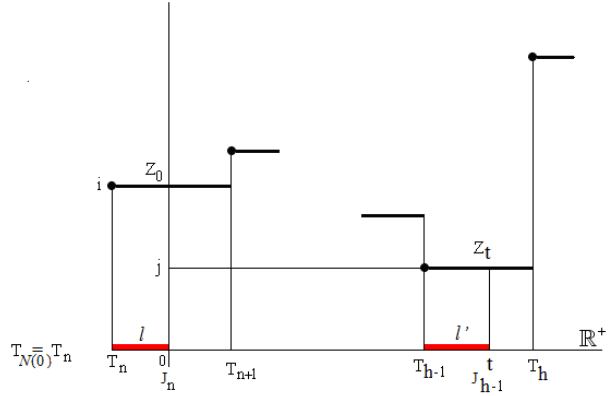}
\caption{Trajectory of a SMC with backward times.}
\label{gh}
\end{figure}

\indent In Figure \ref{gh} we show a trajectory of a SMC. At time $t$ the process $Z(t)$ is in the state $J_{h-1}$ and the last transition occurred at time $T_{h-1}$ then at time $t$ the backward process holds $B(t)=t-T_{h-1}$.

The joint stochastic process $(Z(t),B(t), t\in \N)$ with values in $E\times \N $ is a Markov process, see for example \cite{lim01}. That is:
\begin{displaymath}
\begin{aligned}
& P[Z(T)\!=\!j, B(T)\! \leq \!v'|\sigma(Z(h),B(h)),h\! \leq \!t, Z(t)\!=\!i, B(t)\!=\!v]\\
& =P[Z(T)=j, B(T)\leq v'|Z(t)=i, B(t)=v].
\end{aligned}
\end{displaymath}

The SMC price model proposed by \cite{dami11b} assumes that the value of the asset under study is described by the time varying asset price $S(t)$. The time variable $t\in \{0, 1, \ldots, n \}$ where $n$ is the number of considered unit periods.
 
The return at time $t$ calculated over a time interval of length 1, is defined as
\begin{equation}
Z(t) = \frac{S(t + 1) - S(t)}{S(t)}.
\end{equation}
\indent We assume that $Z(t)$ is a SMC with finite state space $$E=\{-z_{min}\Delta,\ldots,-2\Delta,-\Delta,0,\Delta,2\Delta,\ldots, z_{max}\Delta\}$$
and kernel ${\bf{Q}}=(Q_{ij}(\gamma))$, $\forall i,j\in E$ and $\gamma \in \N $.

\subsection{The Semi-Markov Model with Memory}
In this subsection we propose a generalization of the SMC that is able to represent higher-order dependencies between successive observations of a state variable. One way to increase the memory of the process is by using high-order SMC as defined in \cite{limn03}. Here we propose a more parsimonious model having the objective of defining a new model that appropriately describes empirical regularities of financial time series. To this end we extend the model described above allowing the possibility of reproducing long-term dependence in the square of stock returns. This model was advanced by \cite{dami11a} were a more complete treatment can be found.

\indent Let $(\Omega,\mathbf{F},P)$ be a probability space and consider the stochastic process
$$
J_{-(m+1)},J_{-m},J_{-(m-1)},...,J_{-1},J_{0},J_{1},...
$$
with a finite state space $E=\{1,2,...,S\}$. In our framework the random variable $J_{n}$ describes the price return process at the $n$-th transition.\\
\indent Let us consider the stochastic process
$$
T_{-(m+1)},T_{-m},T_{-(m-1)},...,T_{-1},T_{0},T_{1},...
$$
with values in $\R$. The random variable $T_{n}$ describes the time in which the $n$-th transition of the price return process occurs.\\
\indent Let us consider also the stochastic process
$$
U_{-(m+1)},U_{-m},U_{-(m-1)},...,U_{-1},U_{0},U_{1},...
$$
with values in $\R$. The random variable $U_{n}$ describes the value of the index process at the $n$-th transition.\\
\indent In reference \cite{dami11b} the process $\{U_{n}\}$ was defined as a reward accumulation process linked to the Markov Renewal Process $\{J_{n},T_{n}\}$. In this paper we introduce a different index process $U_{n}^{m}$ that is defined as follows:
\begin{equation}
\label{funcrel}
U_{n}^{m}=\frac{1}{T_{n}-T_{n-(m+1)}}\sum_{k=0}^{m}\int_{T_{n-1-k}}^{T_{n-k}}f(J_{n-1-k},s)ds,
\end{equation}
where $f:E\times \R \rightarrow \R$ is a Borel measurable bounded function and $U_{-(m+1)}^{m},...,U_{0}^{m}$ are known and non-random.\\
\indent The process $U_{n}^{m}$ can be interpreted as a moving average of the accumulated reward process with the function $f$ as a measure of the rate of reward per unit time.\\
\indent The function $f$ depends on the state of the system $J_{n-1-k}$ and on the time $s$.

As an example you can think of the case in which $m=1$ and $f(J_{n},s)=(J_{n})^{2}$. In this simple case we have that:
\begin{equation}
\label{exa}
U_{n}^{1}=\frac{1}{T_{n}-T_{n-2}}\Bigg((J_{n-1})^{2}\cdot(T_{n}-T_{n-1})+(J_{n-2})^{2}\cdot(T_{n-1}-T_{n-2})\Bigg),
\end{equation}
which expresses a moving average of order $m+1=2$ executed on the series of the square of returns with weights given by the fractions
\begin{equation}
\frac{T_{n}-T_{n-1}}{T_{n}-T_{n-2}};\,\,\,\frac{T_{n-1}-T_{n-2}}{T_{n}-T_{n-2}}.
\end{equation}
It should be noted that the order of the moving average is on the number of transitions. As a consequence, the moving average is executed on time windows of variable length.\\
\indent To construct an indexed model we have to specify a dependence structure between the variables. Toward this end we adopt the following assumption:
\begin{equation}
\label{kerne}
\begin{aligned}
& P[J_{n+1}=j,\: T_{n+1}-T_{n}\leq t |\sigma(J_{h},T_{h},U_{h}^{m}),\, h=-m,...,0,...,n, J_{n}=i, U_{n}^{m}=v]\\
& =P[J_{n+1}=j,\: T_{n+1}-T_{n}\leq t |J_{n}=i, U_{n}^{m}=v]:=Q_{ij}^{m}(v;t),
\end{aligned}
\end{equation}
\noindent where $\sigma(J_{h},T_{h},U_{h}^{m}),\, h\leq n$ is the natural filtration of the three-variate process.\\
\indent The matrix of functions ${\bf Q}^{m}(v;t)=(Q_{ij}^{m}(v;t))_{i,j\in E}$ has a fundamental role in the theory we are going to expose. In recognition of its importance, we call it $\emph{indexed semi-Markov}$ $\emph{kernel}$.\\
\indent The joint process $(J_{n},T_{n})$, which is embedded in the indexed semi-Markov kernel, depends on the moving average process $U_{n}^{m}$, the latter acts as a stochastic index. Moreover, the index process $U_{n}^{m}$ depends on $(J_{n},T_{n})$ through the functional relationship $(\ref{funcrel})$.\\
\indent To describe the behavior of our model at whatever time $t$ we need to define additional stochastic processes.\\
\indent Given the three-dimensional process $\{J_{n}, T_{n}, U_{n}^{m}\}$ and the indexed semi-Markov kernel ${\bf Q}^{m}(v;t)$, we define by
\begin{equation}
\label{stocpro}
\begin{aligned}
& N(t)=\sup\{n\in \mathbb{N}: T_{n}\leq t\};\\
& Z(t)=J_{N(t)};\\
& U^{m}(t)=\frac{1}{t-T_{(N(t)-\theta)-m}}\sum_{k=0}^{m}\int_{T_{(N(t)-\theta)-k}}^{t\wedge T_{(N(t)-\theta)+1-k}}f(J_{(N(t)-\theta)-k},s)ds,
\end{aligned}
\end{equation}
where $T_{N(t)}\leq t < T_{N(t)+1}$ and $\theta =1_{\{t=T_{N(t)}\}}$.\\
\indent The stochastic processes defined in $(\ref{stocpro})$ represent the number of transitions up to time $t$, the state of the system (price return) at time $t$ and the value of the index process (moving average of function of price return) up to $t$, respectively. We refer to $Z(t)$ as an indexed semi-Markov process.\\
\indent The process $U^{m}(t)$ is a generalization of the process $U_{n}^{m}$ where time $t$ can be a transition or a waiting time. It is simple to realize that if $\forall m$, if  $t=T_{n}$ we have that $U^{m}(t)=U_{n}^{m}$.\\  
\indent Let 
$$
p_{ij}^{m}(v):= P[J_{n+1}=j|J_{n}=i,U_{n}^{m}=v].
$$ 
be the transition probabilities of the embedded indexed Markov chain. It denotes the probability that the next transition is in state $j$ given that at current time the process entered in state $i$ and the index process is $v$. It is simple to realize that
\begin{equation}
p_{ij}^{m}(v)=\lim_{t\rightarrow \infty}Q_{ij}^{m}(v;t).
\end{equation}
\indent Let $H_{i}^{m}(v;\cdot)$ be the sojourn time cumulative distribution in state $i\in E$:
\begin{equation}
H_{i}^{m}(v;t):= P[ T_{n+1}-T_{n} \leq t |  J_n=i,\, U_{n}^{m}=v ]= \sum_{j\in E}Q_{ij}^{m}(v;t).
\end{equation}
\indent It expresses the probability to make a transition from state $i$ with sojourn time less or equal to $t$ given the indexed process is $v$.\\  \indent The conditional waiting time distribution function $G$ expresses the following probability:
\begin{equation}
\label{G1}
G_{ij}^{m}(v;t):=P[T_{n+1}-T_{n}\leq t \mid J_{n}=i, J_{n+1}=j,U_{n}^{m}=v].
\end{equation}
\indent It is simple to establish that
\begin{eqnarray}
&&G_{ij}^{m}(v;t)=\left\{
                \begin{array}{cl}
                       \ \frac{Q_{ij}^{m}(v;t)}{p_{ij}^{m}(v)}  &\mbox{if $p_{ij}^{m}(v)\neq 0$}\\
                         1  &\mbox{if $p_{ij}^{m}(v)=0$}.\\
                   \end{array}
             \right.
\end{eqnarray}
\indent To properly assess the probabilistic behavior of the system, we introduce the transition probability function:
\begin{equation}
\begin{aligned}
& \phi_{(i_{-(m+1)},i_{-m},...i_{0};j)}^{m}(t_{-(m+1)},t_{-m},...,t_{0};t,V):=\\
& P[Z(t)=j, U^{m}(t)\leq V|J_{0}=i_{0},...,J_{-(m+1)}=i_{-(m+1)},T_{0}=t_{0},...,T_{-(m+1)}=t_{-(m+1)}].
\end{aligned}
\end{equation}
\indent In \cite{dami11a} it was shown that the transition probability function of the ISMC satisfies a renewal-type equation. 

\subsection{The Weighted-Indexed Semi-Markov Model}
\label{fre}
In this subsection we describe a further improvement of the SMC model described above, named Weighted-Indexed Semi-Markov Chain (WISMC) model which allows the possibility of reproducing long-term dependence in the square of stock returns in a very efficient way; this model was proposed and investigated by \cite{dami12}.

\indent  Let us assume that the value of the financial asset under study is described by the time varying asset price $S(t)$. The return at time $t$ calculated over a time interval of lenght $1$ is defined as $\frac{S(t+1)-S(t)}{S(t)}$. The return process changes value in time, then we denote by $\{J_{n}\}_{n\in \N}$ the stochastic process with finite state space $E=\{1,2,...,s\}$ and describing the value of the return process at the $n$-th transition.\\
\indent Let us consider the stochastic process $\{T_{n}\}_{n\in \N}$ with values in $\N$. The random variable $T_{n}$ describes the time in which the $n$-th transition of the price return process occurs.\\
\indent Let us consider also the stochastic process $\{U_{n}^{\lambda}\}_{n\in \N}$ with values in $\R$. The random variable $U_{n}^{\lambda}$ describes the value of the index process at the $n$-th transition.\\
\indent In reference \cite{dami11a} the process $\{U_{n}\}$ was defined as a reward accumulation process linked to the Markov Renewal Process $\{J_{n},T_{n}\}$; in \cite{dami11a} the process $\{U_{n}\}$ was defined as a moving average of the reward process. Here, motivated by the application to financial returns, we consider a more flexible index process defined as follows:
\begin{equation}
\label{funcrela}
U_{n}^{\lambda}=\sum_{k=0}^{n-1}\sum_{a=T_{n-1-k}}^{T_{n-k}-1}f(J_{n-1-k},a,\lambda),
\end{equation}
where $f:E\times \N \times \R \rightarrow \R$ is a Borel measurable bounded function and $U_{0}^{\lambda}$ is known and non-random.\\
\indent The process $U_{n}^{\lambda}$ can be interpreted as an accumulated reward process with the function $f$ as a measure of the weighted rate of reward per unit time. The function $f$ depends on the current time $a$, on the state $J_{n-1-k}$ visited at current time and on the parameter $\lambda$ that represents the weight.\\
In next section a specific functional form of $f$ will be selected in order to produce a real data application.\\
\indent To construct the WISMC model we have to specify a dependence structure between the variables. Toward this end we adopt the following assumption:
\begin{equation}
\label{kernel}
\begin{aligned}
& \mathbb{P}[J_{n+1}=j,\: T_{n+1}-T_{n}\leq t |\sigma(J_{h},T_{h},U_{h}^{\lambda}),\, h=0,...,n, J_{n}=i, U_{n}^{\lambda}=v]\\
& =\mathbb{P}[J_{n+1}=j,\: T_{n+1}-T_{n}\leq t |J_{n}=i, U_{n}^{\lambda}=v]:=Q_{ij}^{\lambda}(v;t),
\end{aligned}
\end{equation}
\noindent where $\sigma(J_{h},T_{h},U_{h}^{\lambda}),\, h\leq n$ is the natural filtration of the three-variate process.\\
\indent The matrix of functions ${\bf Q}^{\lambda}(v;t)=(Q_{ij}^{\lambda}(v;t))_{i,j\in E}$ has a fundamental role in the theory we are going to expose, in recognition of its importance, we call it $\emph{weighted-indexed}$ $\emph{semi-Markov}$ $\emph{kernel}$.\\
\indent The joint process $(J_{n},T_{n})$ depends on the process $U_{n}^{\lambda}$, the latter acts as a stochastic index. Moreover, the index process $U_{n}^{\lambda}$ depends on $(J_{n},T_{n})$ through the functional relationship $(\ref{funcrela})$.\\
\indent Observe that if 
\[
\mathbb{P}[J_{n+1}=j,\: T_{n+1}-T_{n}\leq t |J_{n}=i, U_{n}^{\lambda}=v]=\mathbb{P}[J_{n+1}=j,\: T_{n+1}-T_{n}\leq t |J_{n}=i]
\]
\noindent for all values $v\in \R$ of the index process, then the weigthed indexed semi-Markov kernel degenerates in an ordinary semi-Markov kernel and the WISMC model becomes equivalent to classical semi-Markov chain model as presented for example in \cite{jans06} and \cite{barb08}.\\
\indent The triple of processes $\{J_{n}, T_{n}, U_{n}^{\lambda}\}$ describes the behaviour of the system only in correspondence of the transition times $T_{n}$. To describe the behavior of our model at whatever time $t$ which can be a transition time or a waiting time, we need to define additional stochastic processes.\\
\indent Given the three-dimensional process $\{J_{n}, T_{n}, U_{n}^{\lambda}\}$ and the weighted indexed semi-Markov kernel ${\bf Q}^{\lambda}(v;t)$, we define by
\begin{equation}
\label{stocproc}
\begin{aligned}
& N(t)=\sup\{n\in \mathbb{N}: T_{n}\leq t\};\\
& Z(t)=J_{N(t)};\\
& U^{\lambda}(t)=\sum_{k=0}^{N(t)-1+\theta}\,\,\sum_{a=T_{N(t)+\theta -1-k}}^{(t\wedge T_{N(t)+\theta-k})-1}f(J_{N(t)+\theta-1-k},a,\lambda),
\end{aligned}
\end{equation}
where $\theta =1_{\{t>T_{N(t)}\}}$.\\
\indent The stochastic processes defined in $(\ref{stocproc})$ represent the number of transitions up to time $t$, the state of the system (price return) at time $t$ and the value of the index process (weighted moving average of function of price return) up to $t$, respectively. We refer to $Z(t)$ as a weighted indexed semi-Markov process.\\
\indent The process $U^{\lambda}(t)$ is a generalization of the process $U_{n}^{\lambda}$ where time $t$ can be a transition or a waiting time. It is simple to realize that if $t=T_{n}$ we have that $U^{\lambda}(t)=U_{n}^{\lambda}$.\\  
\indent Let 
$$
p_{ij}^{\lambda}(v):= \mathbb{P}[J_{n+1}=j|J_{n}=i,U_{n}^{\lambda}=v],
$$ 
be the transition probabilities of the embedded indexed Markov chain. It denotes the probability that the next transition is in state $j$ given that at current time the process entered in state $i$ and the index process is equal to $v$. It is simple to realize that
\begin{equation}
p_{ij}^{\lambda}(v)=\lim_{t\rightarrow \infty}Q_{ij}^{\lambda}(v;t).
\end{equation}
\indent Let $H_{i}^{\lambda}(v;\cdot)$ be the sojourn time cumulative distribution in state $i\in E$:
\begin{equation}
H_{i}^{\lambda}(v;t):= \mathbb{P}[ T_{n+1}-T_{n} \leq t |  J_n=i,\, U_{n}^{\lambda}=v ]= \sum_{j\in E}Q_{ij}^{\lambda}(v;t).
\end{equation}
\indent It expresses the probability to make a transition from state $i$ with sojourn time less or equal to $t$ given the indexed process is $v$.\\  \indent The conditional waiting time distribution function $G$ expresses the following probability:
\begin{equation}
\label{G}
G_{ij}^{\lambda}(v;t):=\mathbb{P}[T_{n+1}-T_{n}\leq t \mid J_{n}=i, J_{n+1}=j,U_{n}^{\lambda}=v].
\end{equation}
\indent It is simple to establish that
\begin{eqnarray}
&&G_{ij}^{\lambda}(v;t)=\left\{
                \begin{array}{cl}
                       \ \frac{Q_{ij}^{\lambda}(v;t)}{p_{ij}^{\lambda}(v)}  &\mbox{if $p_{ij}^{\lambda}(v)\neq 0$}\\
                         1  &\mbox{if $p_{ij}^{\lambda}(v)=0$}.\\
                   \end{array}
             \right.
\end{eqnarray}

\section{Empirical Results}

To check the validity of our models we perform a comparison of the behavior of real data returns
and returns generated through Monte Carlo simulations based on the models.
In this section we describe the database of real data used for the analysis, the method used to simulate
synthetic returns time series and, at the end, we compare results from real and simulated data.

\subsection{Database Description}
The data we used to compare results between real data and the SMC model and the ISMC model are tick-by-tick quotes of indexes and stocks downloaded from $www.borsaitaliana.it$ for the period January 2007-December 2010 (4 full years). 
The data have been re-sampled to have 1 minute frequency. Consider a single day (say day $k$ with $1 \le k \le d$)  
where $d$ is number of traded days in the time series. In our case we consider 
four years of trading (from the first of 
January 2007 corresponding to $d=1076$).
The market in Italy fixes the opening price at a
random time in the first minute after 9 am,
continuous trading starts immediately after and ends just before  5.25 pm,
finally the closing price is fixed just after 5.30 pm.
Therefore, let us define $S(t)$ as the price of the last trading
before 9.01.00 am , $S(t+1)$ as the price of the last trading
before 9.02.00 am and so on until
$S(nk)$ as the price of the last trading
before 5.25.00  pm.  
If there are no transactions in the minute,
the price remains unchanged
(even in the case the title is suspended and reopened
in the same day).
Also define
$S(nk+1)$ as the opening price and $S(nk)$ as the closing price.
With this choice $n=507$.
There was a small difference before the 28th of September 2009 since
continuous trading started at 9,05 am, and therefore
prior of that date we have $n=502$.
Finally, if the title has a delay in opening
or it closes in advance (suspended but not reopened),
only the effective trading minutes 
are taken into account. In this case $n$ will be smaller than 507.
The number of returns analyzed is then roughly 508000 for each stock.
We analyzed all the stocks in the FTSEMIB which are the 40 most capitalized
stocks in the Italian stock market.

\begin{figure}
\centering
\includegraphics[width=8cm]{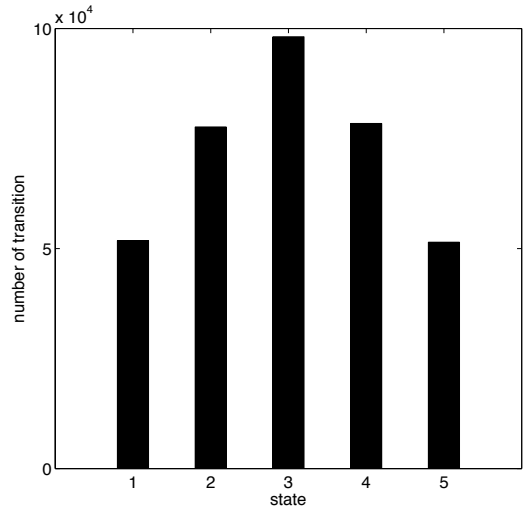}
\caption{Number of transitions for the embedded Markov chain.} \label{tra}
\end{figure}

To be able to model returns as a SMC the state space has to be discretized.
In the example shown we discretized returns into 5 states chosen to be symmetrical with respect to returns equal zero. Returns are in fact already discretized in real data due to the discretization of stock prices which is fixed by each stock exchange and depends on the value of the stock. Just to make an example, in the Italian stock market for stocks with value between 5.0001 and 10 euros the minimum variation is fixed to 0.005 euros (usually called tick). We then tried to remain as much as possible close to this discretization. In Figure \ref{tra} we show an example of the number of transition from state $i$ to all other states for the embedded Markov chain.

For the WISMC model we choose 4 stocks from two databases of tick-by-tick quotes of real stocks from the Italian Stock Exchange (``Borsa Italiana")
and the German Stock Exchange (``Deutsche B\"orse"). The chosen stocks are ENI and FIAT from the Italian database and Allianz and VolksWagen from the German database.The period used goes from January 2007 to December 2010 (4 full years). 
The data have been re-sampled to have 1 minute frequency. The number of returns analyzed is then roughly $500*10^3$ for each stock.

\begin{figure}
\centering
\includegraphics[width=8cm]{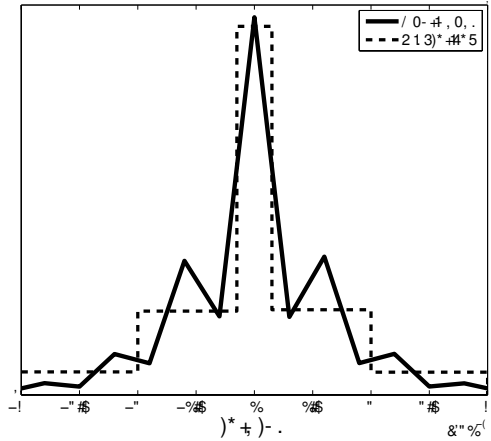}
\caption{Discretization of returns.} \label{discR}
\end{figure}

In these 4 examples, we discretized returns into 5 states chosen to be symmetrical with respect to returns equal zero and to keep the shape of the distribution unchanged. Also in this case, returns are already discretized. In Figure \ref{discR} we show an example of the discretization of the returns of one of the analyzed stocks. 

From the discretized returns we estimated the probabilities ${\bf P}$ and $G_{ij}(t)$ to generate a synthetic trajectory of the described models. 
 For this aim we derive a Monte Carlo algorithm in order to simulate a trajectory of a given model in the time interval $[0, T]$. $T$ is chosen such as the synthetic time series have the same length as the real one.
 The output of the algorithm consists in the successive visited states $\{J_{0}, J_{1},...\}$, the jump times $\{T_{0},T_{1},...\}$  up to the time $T$.The follow algorithm is an example of Monte Carlo simulation for the semi-Markov model:\\
1) Set $n=0$, $J_{0}=i$, $T_{0}=0$, horizon time$=T$;\\
2) Sample $J$ from $p_{J_{n},\cdot}$ and set $J_{n+1}=J(\omega)$;\\
3) Sample $W$ from $G_{J_{n},J_{n+1}}(\cdot)$ and set $T_{n+1}=T_{n}+W(\omega)$;\\
4) If $T_{n+1}\geq T$ stop\\
\indent else set $n=n+1$ and go to 2).

\subsection{Results on the Autocorrelation Function}
A very important feature of stock market data is that, while returns are uncorrelated and show an i.i.d. like behavior, their square or absolute values are long range correlated. It is very important that theoretical models of returns do reproduce this features.   
We then tested our models to check whether it is able to reproduce such behavior. 

We remind the definition of the autocorrelation function: if $Z$ indicates returns, the time lagged $(\tau)$ autocorrelation of the square of returns is defined as 
\begin{equation}
\label{autosquare}
\Sigma(\tau)=\frac{Cov(Z^2(t+\tau),Z^2(t))}{Var(Z^2(t))}
\end{equation}
We estimated $\Sigma(\tau)$ for real data and for returns time series simulated with different models. 
The time lag $\tau$ was made to run from 1 minute up to 100 minutes. Note that to be able to compare results for $\Sigma(\tau)$ each
simulated time series was generated with the same length as real data.

\begin{figure}
\centering
\includegraphics[height=8cm]{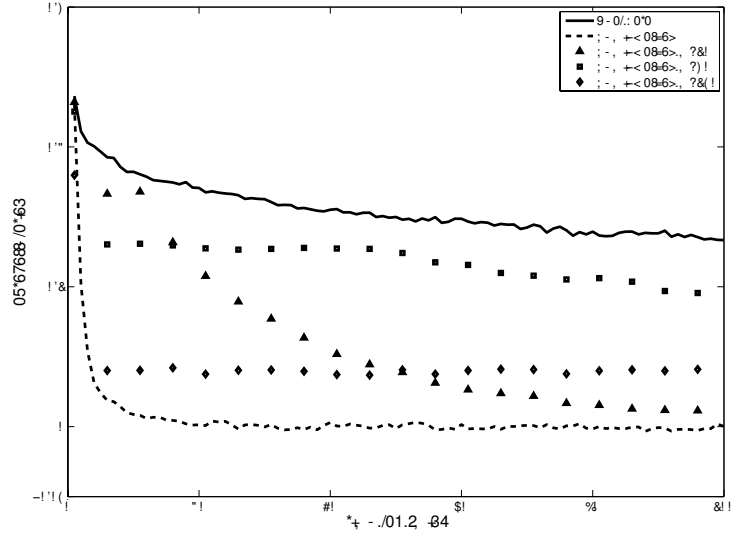}
\caption{Autocorrelation functions of real data (solid line) and of 4 synthetic time series as described in the label.}\label{fig1}
\end{figure}

Results for the indexed semi-Markov model (few values of $m$), for real data and for a semi-Markov model without index are shown in Figure \ref{fig1}.

\begin{figure}
\centering
\includegraphics[height=8cm]{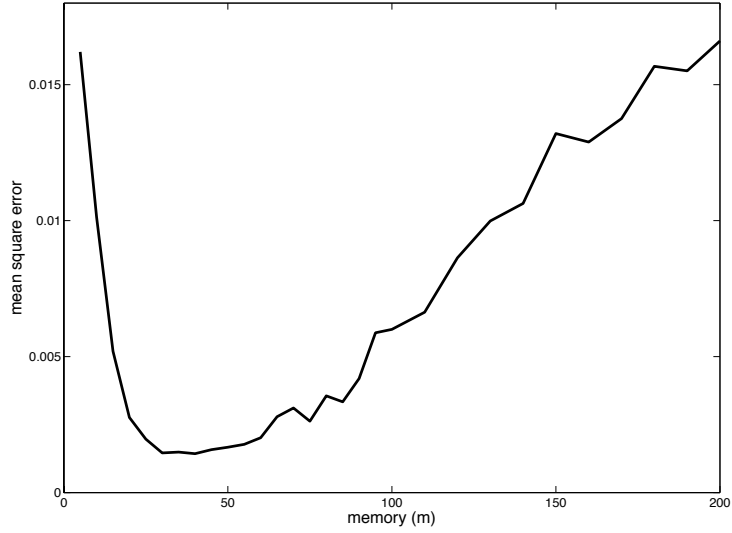}
\caption{Mean square error between autocorrelation function of real data and synthetic data as a function of the memory value $m$.}\label{fig2}
\end{figure}

As expected, real data do show a long range correlation of volatility let us than analyze results for the synthetic time series. The simple semi-Markov model starts at the same value but the persistence is very short and after few time steps the autocorrelation decrease to zero. A very interesting behavior is instead shown by the semi-Markov models with memory index. If a small memory ($m=10$ in the shown example) is used, the autocorrelation is already persistent but again decreases faster than real data. With a longer memory ($m=30$) the autocorrelation remain high for a very  long period and also its value is very close to that of real data. If $m$ is increased further the autocorrelation drops again to small values.
This behavior suggest the existence of an optimal memory $m$. In our opinion one can justify this behavior by saying that short memories are not enough to identify in which volatility status is the market, too long memories mix together different status and then much of the information is lost in the average. 
All this is shown in Figure \ref{fig2} where the mean square error between each autocorrelation function of simulated time series and the autocorrelation function of the real data as a function of $m$ is computed. It can be noticed that there exist an optimal value of the memory $m$ that makes the autocorrelation of simulated data closer to that of real data. 

\begin{figure}
\centering
\includegraphics[width=8cm]{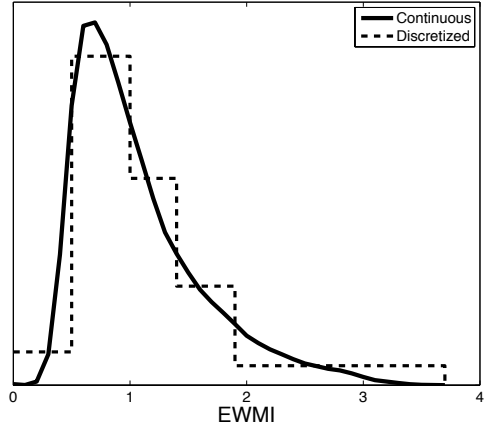}
\caption{Discretization of index values.} \label{discU}
\end{figure}

For what regard the WISMC model, described in the section \ref{fre}, this requires the specification of a function $f$ in the definition of the weighted index $U_{n}^{\lambda}$ in (\ref{funcrela}). 
Let us briefly remind that volatility of real market is long range positively autocorrelated and then clustered in time. This implies that, in the stock market, there are periods of high and low volatility. Motivated by this empirical facts we suppose that also the transition probabilities depends on whether the market is in a high volatility period or in a low one.  In contrast to the indexed semi-Markov model here we decided to use a more appropriate expression for $f$. We use an exponentially weighted moving average (EWMA) of the squares of returns which as the following expression:
 \begin{equation}
\label{funct}
f(J_{n-1-k},a,\lambda)=\frac{\lambda^{T_{n}-a} J_{n-1-k}^2}{\sum_{k=0}^{n-1}\sum_{a=T_{n-1-k}}^{T_{n-k}-1}\lambda^{T_{n}-a}}
\end{equation}
\noindent and consequently the index process becomes
\begin{equation}
\label{ewma}
U_{n}^{\lambda}=\sum_{k=0}^{n-1}\sum_{a=T_{n-1-k}}^{T_{n-k}-1}\Bigg(\frac{\lambda^{T_{n}-a} J_{n-1-k}^2}{\sum_{k=0}^{n-1}\sum_{a=T_{n-1-k}}^{T_{n-k}-1}\lambda^{T_{n}-a}}\Bigg).
\end{equation}
The index $U^\lambda$ was also discretized into 5 states of low, medium low, medium, medium high and high volatility. An example of the discretization used in the analysis is shown in Figure \ref{discU}.

Given the presence of the parameter $\lambda$ in the index function, we tested the autocorrelation behavior as a function of $\lambda$. Note that in the definition of the index variable the EWMA is performed over all the previous square of returns each with its weight. Before summing over all past returns we decided to check whether a better memory time $m$ exists. For this reason we checked our model also against this other parameter.
With this choice formula $(\ref{ewma})$ takes the form:
\begin{equation}
\label{ewma_m}
U_{n}^{\lambda}(m)=\sum_{k=n-m}^{n-1}\sum_{a=T_{n-1-k}}^{T_{n-k}-1}\Bigg(\frac{\lambda^{T_{n}-a} J_{n-1-k}^2}{\sum_{k=n-m}^{n-1}\sum_{a=T_{n-1-k}}^{T_{n-k}-1}\lambda^{T_{n}-a}}\Bigg).
\end{equation}

\begin{figure}
\centering
\includegraphics[height=9cm]{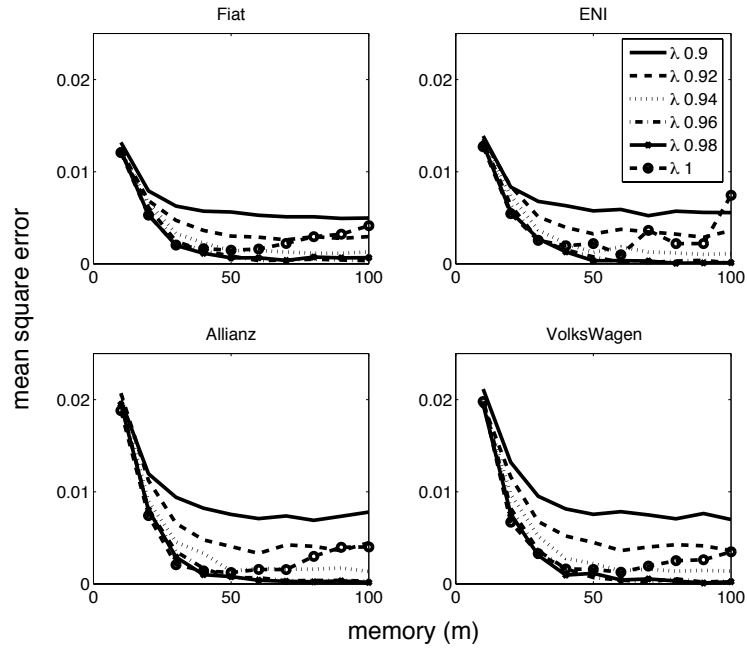}
\caption{Mean square error between autocorrelation functions from real and simulated data as functions of $m$ and for different values of $\lambda$.}\label{fighu}
\end{figure}

\begin{figure}
\centering
\includegraphics[height=9cm]{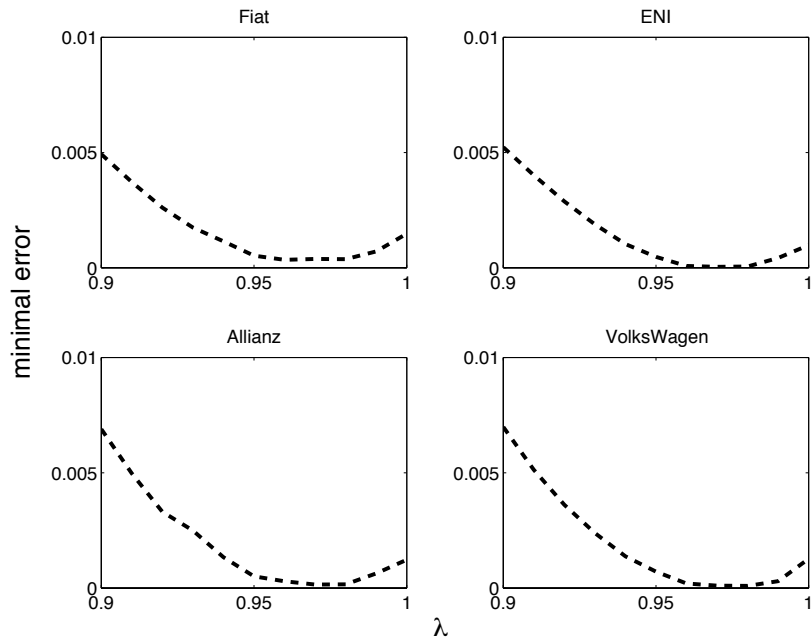}
\caption{Mean square error between autocorrelation functions from real and simulated data as functions of $\lambda$.}\label{figlo}
\end{figure}

In Figure \ref{fighu} we show the mean square error between $\Sigma(\tau)$ obtained from real and simulated returns (using definition $(\ref{ewma_m})$ for the index process) for the four stocks analyzed and for different $m$ and  $\lambda$.
Let us make some considerations on the results shown in Figure \ref{fighu}: $m$ should be chosen as big as possible and then definition $(\ref{ewma})$ is appropriate as far as $\lambda$ is chosen less than $1$, in fact, in this last case definition (\ref{ewma}) becomes equivalent to a moving average without weights and results presented in \cite{dami11a} hold for $m$.
In Figure \ref{figlo} we show again the mean square error but only as a function of the weights $\lambda$ then using definition $(\ref{ewma})$ for the index process. 
We can notice that the behavior is very similar for the different analyzed stocks even if the best value for $\lambda$ is not the same for all of them.
As it is possible to see the best values of $\lambda$ for the stocks Fiat, Eni, Allianz and VolksWagen are $0.96$, $0.97$, $0.97$ and $0.98$, respectively.

\begin{figure}
\centering
\includegraphics[height=8cm]{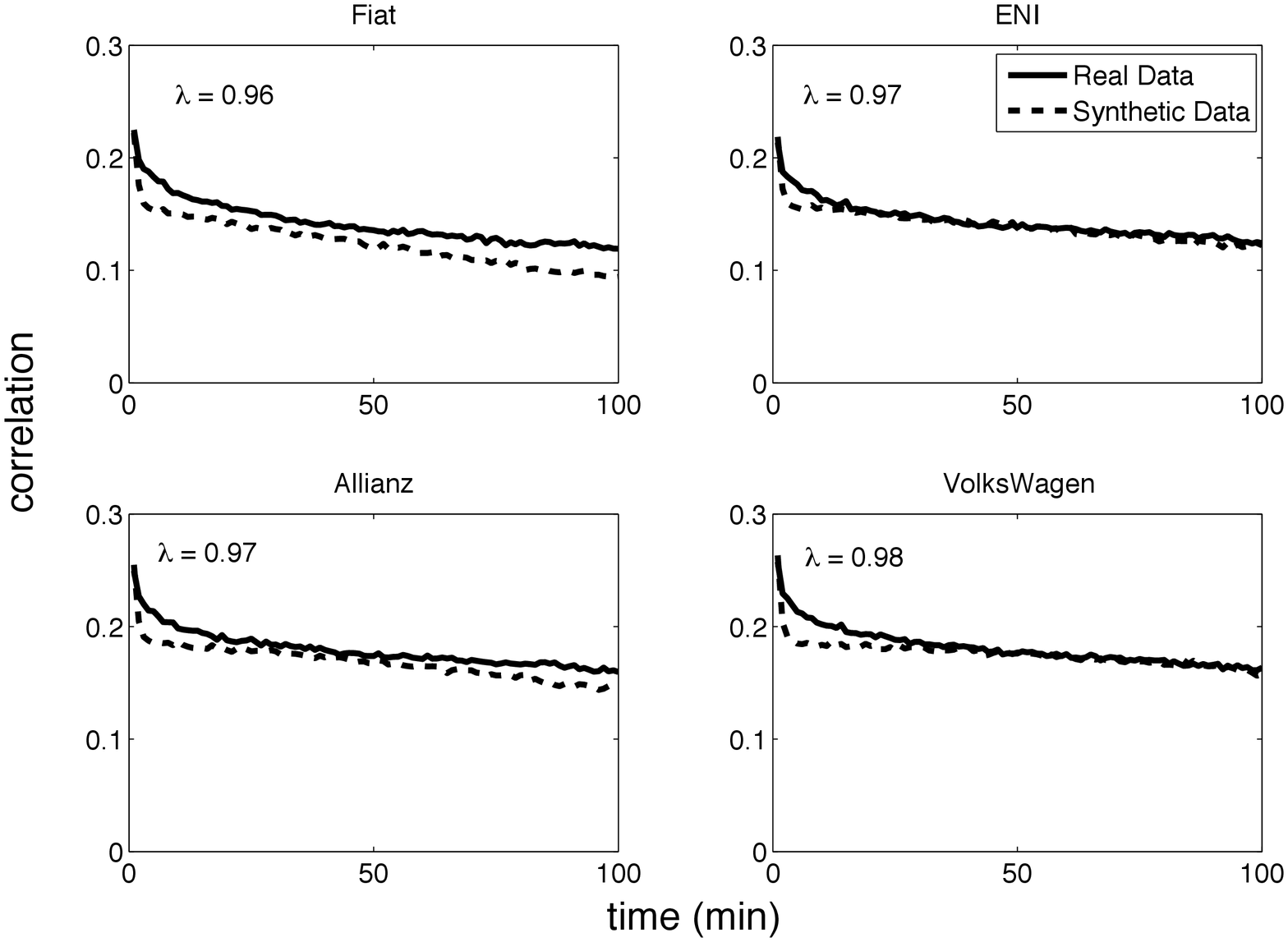}
\caption{Autocorrelation functions of real data (solid line) and synthetic (dashed line) time series for the analyzed stocks.}\label{fig3}
\end{figure}

The comparison between the autocorrelations for the best values of $\lambda$ for each stock and real data is shown in Figure \ref{fig3}.
This figure shows that real and synthetic data have almost the same autocorrelation function for the square of returns.

\section{Concluding Remarks}

We have modeled financial price changes through semi-Markov models. Our work is motivated by the existence in the market of periods of low and high volatility. The simple semi-Markov and the indexed semi-Markov models are able to capture almost all the correlation in the square of returns present in real data. The comparison between these two models shows that the ISMC model reproduce quite well the behavior of the market return thanks to the past volatility used as the memory index. We have shown that the time length of the memory does play a crucial role in reproducing the right autocorrelation persistence, indicating the existence of an optimal value. 

The results presented for the WISMC model, instead, show that if the past volatility is used as an exponentially weighted index, the model is able to reproduce more correctly, than the ISMC model, the behavior of market returns. The returns generated by the model are uncorrelated while the square of returns present a long range correlation very similar to that of real data.

We have also shown, only for the WISMC model, by analyzing different stocks from different markets (Italian and German), that results do not depend on the particular stock chosen for the analysis even if the value of the weights may depends on the stock.

We stress that out models are very different from those of the ARCH/GARCH family. We do not model directly the volatility as a correlated process. We model returns and by considering the semi-Markov kernel and the semi-Markov kernel conditioned by a memory index and a weighted index the volatility correlation comes out freely.


%

\bigskip

MA



\begin{thebibliography}{99}


\bibitem{bar04} V. Barbu, M. Boussemart, N. Limnios. \textit{Commun. Stat. Theory Method} 33 (2004) 2833.

\bibitem{barb08} Barbu, V. and Limnios, N. (2008). {\em Semi-Markov Chains and Hidden Semi-Markov Models Toward Applications}. Springer-Verlag New York Inc.

\bibitem{dami11c} G. D'Amico, {\it Age-usage semi-Markov models}, \textit{Applied Mathematical Modelling}, { 35}, (2011), 4354-4366.

\bibitem{dam05} G. D'Amico, J. Janssen, R. Manca. { Homogeneous semi-Markov reliability
models for credit risk management}, \textit{Decisions in Economics and Finance}, { 28}, (2005), 79-93.

\bibitem{dam09} G. D'Amico, J. Janssen, R. Manca. { European and American options: The semi-Markov case}, \textit{Physica A}, { 388}, (2009), 3181-3194.

\bibitem{dam10} G. D'Amico, J. Janssen, R. Manca. \textit{Methodol Comput Appl Probab.} 12(2) (2010) 215.

\bibitem{dami11b} G. D'Amico, F. Petroni. {\it A semi-Markov model for price returns}, \textit{Physica A} 2012, DOI: 10.1016/j.physa.2012.05.040.

\bibitem{dami11a} G. D'Amico, F. Petroni. {\it A semi-Markov model with memory for price changes}, \textit{Journal of Statistical Mechanics: Theory and Experiment}, P12009, 2011

\bibitem{dami12} G. D'Amico, F. Petroni. { Weighted-indexed semi-Markov models for modeling financial returns." \textit{Journal of Statistical Mechanics: Theory and Experiment}} (2012) P07015.

\bibitem{dami13} G. D'Amico, F. Petroni, F. Prattico. { First and second order semi-Markov chains for wind speed modeling}, \textit{Physica A}, { 392}, (2013), 1194-1201.

\bibitem{dami13bis} G. D'Amico, F. Petroni, F. Prattico. { Reliability Measures of Second-Order Semi-Markov Chain Applied to Wind Energy Production}, \textit{Journal of Renewable Energy}, { 2013}, Article ID 368940, 6 pages.

\bibitem{dami13tris} G. D'Amico, F. Petroni, F. Prattico. { Wind speed modeled as an indexed semi-Markov process}, \textit{Environmetrics}, Article first published online: 24 MAY 2013, DOI: 10.1002/env.2215.

\bibitem{gui97} D.M. Guillaume, M.M. Dacorogna, R.R. DavVe, J.A. MWuller, R.B. Olsen, O.V. Pictet. { From the birds eye to the microscope:
A survey of new stylized facts of the intra-daily foreign exchange markets}, \textit{Finance and Stochastics} { 1}, (1997), 95-129.

\bibitem{jans06} J. Janssen and R. Manca, {\em Applied semi-Markov processes,} Springer, New York, 2006.

\bibitem{jen04} M. H. Jensen, A. Johansen, F. Petroni, I. Simonsen. { Inverse Statistics in the Foreign Exchange Market}. \textit{Physica A} 340, (2004) 678.

\bibitem{lim01} N.Limnios, G. Opri\c{c}an, \textit{Semi-Markov Processes and Reliability Modeling}, Birkh\"{a}user, Boston. (2001).

\bibitem{limn03} N. Limnios, G. Opri\c{s}an. { An introduction to Semi-Markov Processes with Application to Reliability}, In D.N. Shanbhag and C.R. Rao, eds., \textit{Handbook of Statistics}, { 21}, (2003), 515-556.

\bibitem{mai00} F. Mainardi, M. Raberto, R. Gorenflo, E. Scalas.{\it Fractional calculus and continuous-time finance II: The waiting-time distribution}, \textit{Physica A}, { 287}, (2000), 468-481.

\bibitem{pet03} F. Petroni, M. Serva. {\it Spot foreign exchange market and time series}, \textit{The European Physical Journal B} { 34}, (2003), 495-500.

\bibitem{rab02} M. Raberto, E. Scalas, F. Mainardi. {\it Waiting-times and returns in high-frequency financial data: An empirical study}, \textit{Physica A} { 314}, (2002) 749-755.

\bibitem{sil04} D.S. Silvestrov, F. Stenberg. {\it A pricing process with stochastic volatility controlled by a semi-Markov process}, \textit{Communications in Statistics: Theory and Methods}, { 33}, (2004), 591-608.

\bibitem{sim02} I. Simonsen. M.H. Jensen, A. Johansen, { Optimal Investment Horizons}, \textit{Eur. Phys. J.} 27 (2002) 583.


\end{thebibliography}
\end{document}